\documentclass[a4paper]{article}

\usepackage{INTERSPEECH2020}
\newcommand{\bftab}{\fontseries{b}\selectfont}

\usepackage{algorithm}
\usepackage[noend]{algpseudocode}
\usepackage{hyperref}
\usepackage[abs]{overpic}
\makeatletter
\def\BState{\State\hskip-\ALG@thistlm}
\makeatother

\title{Quantization of Acoustic Model Parameters in Automatic Speech Recognition Framework }
\name{Amrutha Prasad, Petr Motlicek, Srikanth Madikeri, Rudolf Braun}
\address{
  Idiap Research Institute, Martigny Switzerland
}
\email{\{aprasad, petr.motlicek, srikanth.madikeri, rudolf.braun\}@idiap.ch }

\begin{document}

\maketitle
\begin{abstract}
  State-of-the-art hybrid automatic speech recognition (ASR) system exploits deep neural network (DNN) based acoustic models (AM) trained with Lattice Free-Maximum Mutual Information (LF-MMI) criterion and n-gram language models. The AMs typically have millions of parameters  and require significant parameter reduction to operate on embedded devices. 
  
  The impact of parameter quantization on the overall word recognition performance is studied in this paper. Following approaches are presented: (i) AM trained in Kaldi framework with conventional factorized TDNN (TDNN-F) architecture, (ii) the TDNN AM built in Kaldi loaded into the PyTorch toolkit using a C++ wrapper for post-training quantization,
(iii) quantization-aware training in PyTorch for Kaldi TDNN model, (iv) quantization-aware training in Kaldi. Results obtained on standard Librispeech setup provide an interesting overview of recognition accuracy w.r.t. applied quantization scheme.

\end{abstract}

\section{Introduction}

 Deep Neural Networks (DNN) help learn multiple levels of representation of data in order to model complex relationships among them. Conventional acoustic models (AM) used in ASR frameworks are trained with neural network architectures such as Convolutional Neural Networks (CNN)~\cite{lecun1995convolutional}, Recurrent Neural Networks (RNNs)~\cite{graves2013speech}, Time-delay Neural Networks (TDNN)~\cite{peddinti2015time}.
 The Kaldi~\cite{povey2011kaldi} toolkit provides state-of-the-art techniques to train acoustic models, specifically Lattice-Free Maximum Mutual Information (LF-MMI)~\cite{povey2016purely}.
 Since such models often have millions of parameters, they become impractical for use in
  embedded devices with limited memory and computational power (e.g. Raspberry Pi). 
  To embed ASR systems on such devices, the footprint of the ASR system needs to be significantly reduced.
  One simple solution is to train a model with fewer parameters.
  However, reducing the model size usually affects the performance of the system negatively. 
 
 Previous research has shown several possible alternative approaches. Quantizing the model parameters
 from floating point values to integers is one popular approach.
In~\cite{jacob2018quantization}, quantization methods are studied for CNN architectures in image classification and other computer vision problems. Results show that quantizing such models reduces the model size significantly without any impact on performance. 
Another approach is to use teacher-student training to first train a larger model that is optimized for performance and use its output to train a smaller model.
Alternatively, in models such as~\cite{povey2018semi} parameter reduction is integrated as a part of training.
In this paper, we study the effect of quantizing the parameters of an AM trained with LF-MMI (with Kaldi) 
with a focus on deploying it on embedded devices with low computational resources (especially, memory).
We present the impact on the performance of the ASR system when the AM is quantized at different levels: from 32-bit floating point (float32) to 8-bit (int8) or 16-bit (16-bit) integers. 
We compare the results to other techniques typically used in parameter reduction for automatic speech recognition models.
We believe that results obtained from our study have not been presented in literature, and can be of interest for researchers experimenting with interfacing Kaldi and PyTorch~\cite{paszke2019pytorch} tools for ASR tasks.
 
 The rest of the paper is organized as follows. Section 2 describes briefly the current techniques used in parameter reduction of a model. This is followed by an overview of the quantization techniques and their application  to AM trained with Kaldi toolkit (Section 3). Section 4 presents the experiments and the results. finally, the conclusion is provided in Section 5.

\section{Related work}

Parameter reduction aims at reducing the footprint of the neural network while preserving the useful information of the network required for its decision process. For DNN-based acoustic models this can be achieved in many ways:

\begin{itemize}
    \item Teacher-student approach to reduce the number of layers in the student neural network~\cite{wong2016sequence}. 
    \item Reduce the size of the layers used in training the neural network through matrix factorization~\cite{keith2018optimizing}.
    \item Reduce the hidden layer dimension (e.g. from 1024 to 512 in each layer of the neural network). 
    \item Reduce the number of hidden layers used in the network.
    \item Quantization of model parameters (e.g. from 32 bit floating precision to 16 bit floating precision)~\cite{jacob2018quantization,krishnamoorthi2018quantizing}. 
\end{itemize}

Single Value Decomposition (SVD) can be applied to the trained models to factorize the learned weight matrix as a product of two smaller factors. SVD then discards the smaller singular values followed by fine tuning of the network parameters to obtain a parameter-reduced model~\cite{xue2013restructuring}. Results have shown that the model size can be reduced by 80\% without loss in the recognition performance.

Another approach to enforce parameter reduction while training a neural network AM is by applying low-rank factorized layers~\cite{povey2018semi}. In semi-orthogonal factorization, the parameter matrix $M$ is factorized as a product of two matrices $A$ and $B$, where $B$ is a semi-orthogonal matrix and $A$ has a smaller “interior” (i.e. rank) than that of $M$. This technique enables training a smaller network from scratch instead of using a pre-trained network for parameter reduction. The LF-MMI training also provides a stable training procedure for semi-orthogonalized matrices. 
These matrices have been studied with TDNN-F (a variant of TDNN with residual connections) and the results have shown that the model size is reduced by 60\%.


\section{Quantization}
Quantization of model parameters is a well-understood technique to reduce a neural-network's footprint. Deep learning frameworks such as Pytorch~\cite{paszke2019pytorch} and Tensorflow~\cite{abadi2016tensorflow} support many operations with respect to model quantization.
However, its application to AM trained with LF-MMI criterion using Kaldi toolkit is not straightforward
as standard implementations do not support quantization-related operations out-of-the-box.
Our focus in this work is to provide such support by integrating state-of-the-art acoustic
modelling technique available in Kaldi with quantization-related operations in Pytorch.
The following subsections explain the standard quantization process for DNNs and how it is applied to the AMs.

\subsection{Overview of quantization process}

Quantization is a process of mapping a set of real valued inputs to a discrete valued outputs. Commonly used quantization types are 16 bits and 8 bits. 
Quantizing model parameters typically involves decreasing the number of bits used to represent the parameters.
Prior to this process, the model may have been trained with IEEE float32 or float64.
A model size can be reduced by a factor of 2 (with 16 bits quantization) and by a factor of 4 (with 8 bits quantization)
if the original model uses float32 representation. 

In addition to the quantization types, there are different quantization modes such as symmetric and asymmetric quantization. As mentioned earlier, a real valued variable $x$ in the range of $(x_{min}, x_{max})$ is quantized to a range $(q_{\mathrm{min}}, q_{\mathrm{max}})$. In symmetric quantization, the range $(q_{\mathrm{min}}, q_{\mathrm{max}})$ corresponds to $(\frac{-N_{levels}}{2}, \frac{N_{levels}}{2} -1)$. In asymmetric quantization the quantization range is $(0, \frac{N_{levels}}{2} -1)$. In the aforementioned intervals, $N_{levels} = 2^{16} = 65536 $ for 16 bit quantization and $N_{levels} = 2^8 = 256$ for 8 bit quantization. 

A real value $r$, can be expressed as an integer $q$ given a scale $S$ and zero-point $Z$~\cite{jacob2018quantization}: 
\begin{equation}
  r = S * (q-Z).
  \label{eq1}
\end{equation}
In the above equation, scale $S$ specifies the step size required to map the floating point to integer and an integer zero-point represents the floating point zero~\cite{jacob2018quantization}.

Given the minimum and maximum of a vector $x$ and the range of the quantization scheme, scale and zero-point is computed as below~\cite{krishnamoorthi2018quantizing}:
\begin{equation}
    S = \frac{x_{max} - x_{min}}{q_{\mathrm{max}} - q_{\mathrm{min}}}
    \label{eq2}
\end{equation}

\begin{equation}
    Z = q_{\mathrm{min}} - \frac{x_{min}}{scale}.
    \label{eq3}
\end{equation}
As mentioned in~\cite{jacob2018quantization}, for 8 bit integer quantization the values never reach -128 and hence we use $q_{\mathrm{min}} = -127$ and $q_{\mathrm{max}} = 127$.

\begin{figure}[t]
    \centering
    
    \includegraphics[width=\textwidth, height=8cm, clip, keepaspectratio, viewport=700 250 1200 1000]{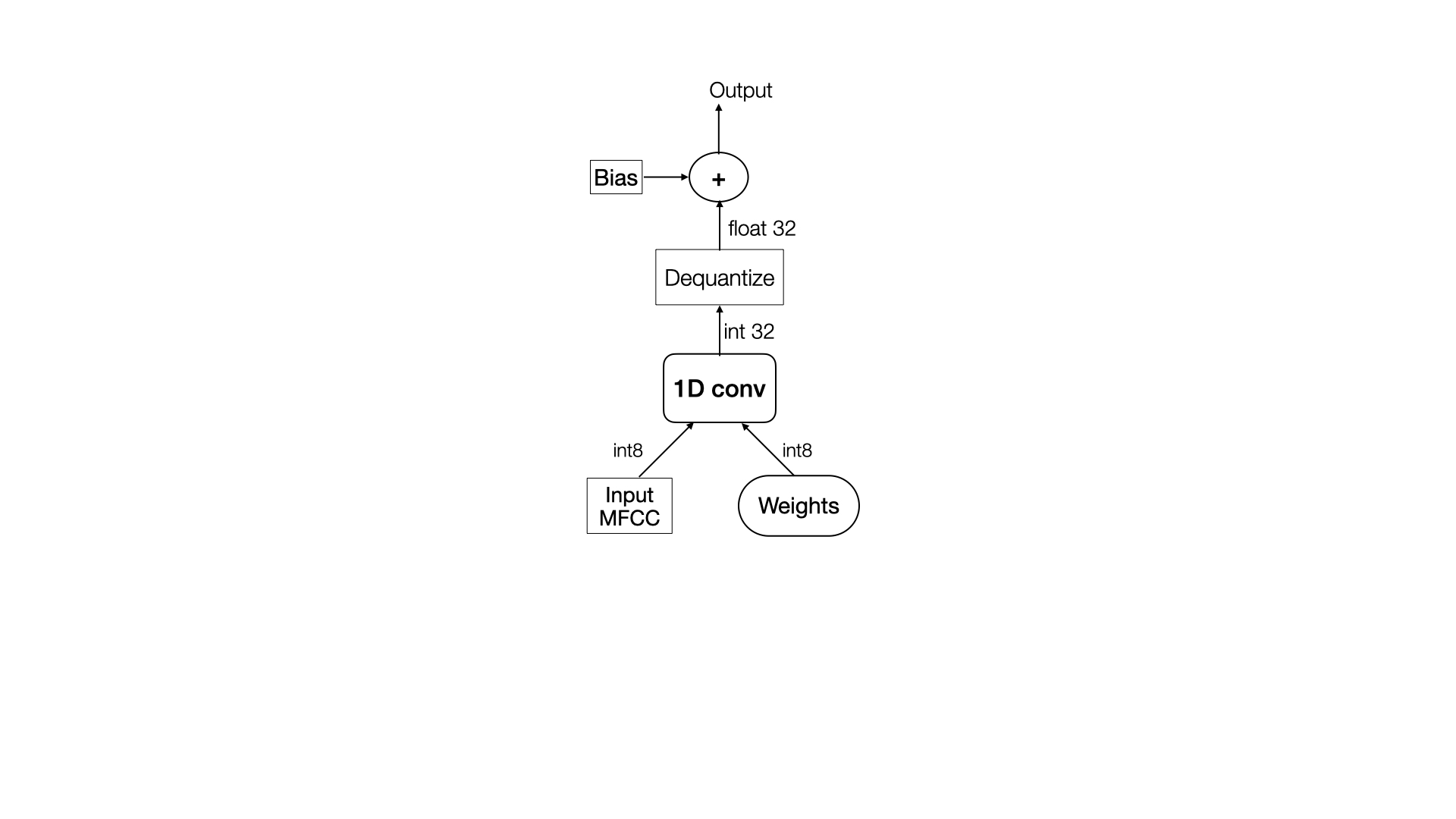}
    \vspace{-12mm}
    \caption{Block diagram of integer arithmetic inference with quantization of weights and activations. Input activations and weights are represented as 8-bit integers according to equation 1. The 1D convolution involves integer inputs and a 32-bit integer accumulator. The output of the convolution is mapped back to floating point and added with the bias. }
    \vspace{-3mm}
\end{figure}

\subsection{Quantization application}
We implement the quantization algorithms in PyTorch as it provides better support than Kaldi for ``int8", ``int8" and ``int16" types.
The aim of our work is to port models trained in Kaldi to be functional in embedded systems. 
There already exist tools such as Pykaldi~\cite{can2018pykaldi} that help users to load Kaldi acoustic models for inference in PyTorch. However, they do not allow access to  the model parameters by default. To support this work, we implemented a C++ wrapper~\cite{madikeri2020pkwrap} that allows access to  the model parameters and input MFCC features as PyTorch tensors. The wrapper also allows us to write the models and ark (archive) files back to Kaldi format. 

Once the model is loaded as a tensor, we perform (i) post-training quantization, (ii) quantization aware training (QAT), (iii) QAT in Kaldi.

\subsubsection{Post-training quantization}
As the name suggests, quantization is applied to a model after it has been trained in Kaldi. We consider two options (a) weights only quantization, (b) quantization of weights and activations. 
In the former, only the weights of the neural network model are quantized. This approach is useful when only the model size needs to be reduced and the inference is carried out in floating-point precision. In our experiments, the weights of the AM are quantized in PyTorch and the inference is carried out in Kaldi. 

In order to reduce the model from 32 bit precision to 8 bit precision, both the weights and activations must be quantized. Activations are quantized with the use of a calibration set to estimate the dynamic range of the activations. Our network architecture consists of TDNN layer followed by ReLu and Batchnorm layers. In our experiments, we quantize only the weights and input activations to the TDNN layer as shown in figure 1 (i.e., the integer arithmetic is applied only to the 1D convolution). Floating point operations are used in ReLu and Batchnorm layers in order to simplify the implementation, as the main focus of this paper is to only study the impact of quantization on AM weights and activations. The conventional word-recognition lattices are then generated by a Kaldi decoder (i.e. performance in Kaldi) with the use of PyTorch generated likelihoods. 

The quantization of weights and activations is done in two ways: (1) custom implementation of quantization and inference with integer arithmetic, and (2) Pytorch implementation.

In our custom implementation, both weights and activations are quantized to ``int8" using Equation~\ref{eq1} and integer arithmetic of the 1d convolution is performed according to the Equation~\ref{eq:int-mm} provided by~\cite{jacob2018quantization}:
\begin{equation}
    q_3^{(i,k)}  = Z_3 + M \left( NZ_1Z_2 - Z_1 a_2^{(k)} - Z_2 a_1^{(i)} +  \sum_{j=1}^N  q_1^{(i,j)} q_2^{(j,k)} \right),
    \label{eq:int-mm}
\end{equation}
where 
$ a_2^{(k)} = \sum_{j=1}^N q_2^{(j,k)}$, $a_1^{(k)} = \sum_{j=1}^N q_1^{(i,j)}$,
$Z_{r}$ represents the zero-point of the weights, activations and the output product,
$q_{r}$ represents the quantized (``int8" representation) weights, activations and the output product and $N$ represents the hidden layer dimension.

Equation~\ref{eq:int-mm} is used to multiply the quantized weights and activations in 1D convolution. The computation of $a_2^{(k)}$ and $a_1^{(i)}$ collectively takes $2N^2$ additions and the last term in equation~\ref{eq:int-mm} takes $2N^3$ operations.

As in our experiments we only perform integer arithmetic of weights and activations, Equation~\ref{eq:int-mm} becomes
\begin{equation}
    r_3^{(i,k)}  = M \left( NZ_1Z_2 - Z_1 a_2^{(k)} - Z_2 a_1^{(i)} +  \sum_{j=1}^N  q_1^{(i,j)} q_2^{(j,k)} \right),
    \label{eq:int-mm1}
\end{equation}
with $M = S_1S_2$. The above equation gives the floating-point value which is then added with the floating-point bias vector. 
This method also allows to experiment with different combination of quantization of weights and activations. 


With PyTorch's builtin quantization functions, weights are quantized to ``int8" (using $torch.nn.quantized.Linear$) and activations to ``uint8" ($torch.quantize\_per\_tensor$). 
The output is then de-quantized for further operations such as ReLu and Batchnorm as mentioned earlier.


\begin{table*}[th]
  \caption{Comparing parameter reduction techniques for monophone and biphone based TDNN acoustic model: Quantization (bits), Weight Quantization (WQ), Activation Quantization (AQ), Number of parameters, model size and Word-Error Rate (WER) [in \%] with the small LM used in decoding. ``post quant" refers to custom implementation of quantization and inference.  }
  \label{tab:quant-m}
  \centering
  \begin{tabular}{ @{}lccccccc@{}}
    \toprule
    Model & Quantization (bits) & WQ & AQ & Params & Size & \multicolumn{2}{c}{WER \%} \\
    \midrule
     & & & & & & Monophone & Biphone \\
    \midrule
    Baseline - no quant & - & No & No & 7.9M  & 1x & 8.1 & 6.3 \\
    \midrule
    \multicolumn{8}{c}{\textit{Post-training quantization}} \\
    \midrule
    16 bit   & 16  & Yes & No & 7.9M & 0.5x & 10.7 & 10.0 \\
    8 bit   & 8  & Yes & No & 7.9M & 0.25x & 10.7 & 11.4 \\
    8 bit (custom)  & 8  & Yes & Yes & 7.9M & 0.25x & \bftab 8.4 & \bftab 7.1 \\
    8 bit (PyTorch built-in)  & 8  & Yes & Yes & 7.9M & 0.25x & 10.4 & 7.5 \\
    \midrule
    \multicolumn{8}{c}{\textit{Quantization aware training}} \\
    \midrule
    Kaldi QAT   & 8 & Yes & Yes & 7.9M & 0.25x & 10.1 & 7.3 \\
    PyTorch QAT & 8  & Yes & Yes & 7.9M & 0.25x & \bftab 8.6  & \bftab 6.9 \\
    \midrule
    TDNN-F  & - & No & No & 3.1M & 0.4x & 10.8 & 8.3\\
    \bottomrule
  \end{tabular}
  
\end{table*}

\subsubsection{Quantization aware training (QAT)}
In this method, during training, weights and activations to the TDNN are ``fake quantized" during the forward and backward pass. This implies that the float values are approximated to mimic the ``int8" values but the floating arithmetic will still be used for computations. The quantization errors are therefore introduced as part of the training loss and the optimizer tries to minimize it during the training. This yields better recognition accuracy compared to the quantization of weights and activations, or post quantization fine-tuning. 

In our experiments, we load the AM trained in Kaldi to Pytorch using pkwrap~\cite{madikeri2020pkwrap}, apply the PyTorch quantization aware training functions and train the model for one epoch. 

\subsubsection{QAT with Kaldi }
A custom Kaldi extension which supports QAT is used in our experiments (note that Kaldi does not support QAT out-of-the-box).
This method follows the usual principle of doing fake quantization by quantizing and dequantizing weights and activations during training. QAT is activated only during the final iterations of the last epoch of training.

    


\section{Experiments}

In all our experiments to parameter reduction, we applied the  TDNN-based acoustic models trained with Kaldi  toolkit (i.e. nnet3 model architecture). AMs are trained with the LF-MMI training framework, considered to produce state-of-the-art performance for hybrid ASR systems. 
In this paper, we not only consider conventional biphone AM, but also a monophone based AM. 
In the former case, the output layer consists of senones obtained from clustering of context-dependent phones.
In the latter case, the output layer consists of only monophone outputs, which can be considered as yet another approach to reduce the computational complexity of ASR systems~\cite{hadian2018end}.
The biphone-based AM uses position-dependent phones which produces a total of 346 phones including the silence and noise phones. The monophone-based AM uses position-independent phones which comprises of 41 phones. The output of the biphone-based AM produces 5984 states while the monophone-based AM produces 41 states. 

\subsection{Experimental setup}
The AMs trained use conventional high-resolution MFCC features with speed perturbed data. We did not include i-vectors. The TDNN and TDNN-F models are composed of  7 layers with the hidden layer dimension of 625.


Our experiments are performed on Librispeech~\cite{panayotov2015librispeech}, a corpus of approximately 1000 hours of 16\,kHz read English speech from the LibriVox project. The LibriVox project is responsible for the creation of approximately 8000 public domain audio books, the majority of which are in English. Most of the recordings are based on texts from Project Gutenberg2, also in the public domain.
The AMs are trained with 960h of Librispeech~\cite{panayotov2015librispeech} data. The LM is also trained on Librispeech which is available for download from the web. The LM used is a 3-gram model  which is pruned with a threshold of $3e-7$. 

The quantization is performed in PyTorch. Quantization experiments are carried out for 16 bit and 8 bit integers in symmetric mode. As discussed in Section 3, the model and the features from Kaldi are loaded as PyTorch tensors with the help of the C++ wrapper~\cite{madikeri2020pkwrap}. 

The word recognition performance for all experiments is evaluated on the Librispeech test-clean set.

\subsection{Parameter reduction experiments}
We compare floating-point vs. integer arithmetic inference for TDNN model with different quantization types (16-bit and 8-bit integer) and different quantization schemes, as discussed in Section 3.2. We also compare the quantization technique with the low-rank matrix factorization technique used during the training of the model.


\subsubsection{Post-training quantization}
(i) Weight only quantization: From Table~\ref{tab:quant-m}, 
quantization reduces the model size by 50\%  for 16 bits and by 75\% for 8 bits but affects ASR performance. The impact on the WER for the monophone model is comparatively low (2.6\% absolute compared to baseline (from 8.1 to 10.7\%)) while the WER drops quite significantly for the biphone model (i.e. $\approx4-5\%$).  

\noindent(ii) Custom implementation of quantization and inference: 
Experiments on both monophone and biphone models show that the model size is reduced by 75\% without any impact on the performance of the ASR system. This is due to the fact that both weights and activations are quantized to the same datatype (``int8") which will have the same range.

\noindent(iii) Quantization and inference with PyTorch functions:
Results on monophone model in Table~\ref{tab:quant-m} show that using this method has a loss in the recognition performance of 2\% (absolute) compared to the previous method (from 8.4 to 10.4\% WER). This is due to the fact that there are fewer outputs in this system. However, this method doesn't show any impact on the performance for the biphone based AM.

\subsubsection{QAT }
Experiments show that QAT implemented in PyTorch reduces quantization error and improves performance of the system for both monophone and biphone systems compared to the post-training quantization with PyTorch functions.
The Kaldi QAT reduces the quantization error and improves performance of the system by 0.6\% (absolute) compared to the weight only quantization (from 10.7\% to 10.1\%) but the loss in the performance compared to baseline is still 2\% (absolute). The advantage of using Kaldi QAT is that switching between Kaldi and PyTorch is not required. 

finally, we also compare the result with a TDNN-F model containing the same number of layers as the baseline TDNN. Model size is reduced by 60\% with a loss in the recognition performance of 2.7\% compared to the baseline TDNN (from 8.1 to 10.8\% WER).
In general, TDNN-F tends to achieve better performance with deeper architectures.


\section{Conclusions}
We presented a study that shows the effect of quantizing the acoustic model parameters in ASR. The experimental results reveal that the parameter-quantization can reduce the model size significantly while preserving the word recognition performance. 
TDNN-F models provide a better performance when the number of layers is higher than for the TDNN models. Quantization of the acoustic models can be further explored through (i) fusing the TDNN, ReLu and Batchnorm layers, (ii) Apply these methods to TDNN-F models. 

The quantization experiments are conducted in PyTorch, while the acoustic models are developed using the popular Kaldi toolkit. The code for post-training quantization and QAT implemented in Pytorch will be offered to other researchers through Github\footnote{\url{ https://github.com/idiap/pkwrap/tree/load_kaldi_models/egs/librispeech/quant}}.

7
\section{Acknowledgements}
This work was partially supported by the CTI Project ``SHAPED: Speech Hybrid Analytics Platform for consumer and Enterprise Devices", as well as by H2020 SESAR Exploratory EC project HAAWAII (grant agreement No. 884287)\footnote{\url{https://www.haawaii.de/}}.  We wish to acknowledge Arash Salarian for providing us with valuable insights and suggestions regarding quantization. 


\bibliographystyle{IEEEtran}

\bibliography{main}

\begin{thebibliography}{10}
\providecommand{\url}[1]{#1}
\csname url@samestyle\endcsname
\providecommand{\newblock}{\relax}
\providecommand{\bibinfo}[2]{#2}
\providecommand{\BIBentrySTDinterwordspacing}{\spaceskip=0pt\relax}
\providecommand{\BIBentryALTinterwordstretchfactor}{4}
\providecommand{\BIBentryALTinterwordspacing}{\spaceskip=\fontdimen2\font plus
\BIBentryALTinterwordstretchfactor\fontdimen3\font minus
  \fontdimen4\font\relax}
\providecommand{\BIBforeignlanguage}[2]{{%
\expandafter\ifx\csname l@#1\endcsname\relax
\typeout{** WARNING: IEEEtran.bst: No hyphenation pattern has been}%
\typeout{** loaded for the language `#1'. Using the pattern for}%
\typeout{** the default language instead.}%
\else
\language=\csname l@#1\endcsname
\fi
#2}}
\providecommand{\BIBdecl}{\relax}
\BIBdecl

\bibitem{lecun1995convolutional}
Y.~LeCun, Y.~Bengio \emph{et~al.}, ``Convolutional networks for images, speech,
  and time series,'' \emph{The handbook of brain theory and neural networks},
  vol. 3361, no.~10, p. 1995, 1995.

\bibitem{graves2013speech}
A.~Graves, A.-r. Mohamed, and G.~Hinton, ``Speech recognition with deep
  recurrent neural networks,'' in \emph{2013 IEEE international conference on
  acoustics, speech and signal processing}.\hskip 1em plus 0.5em minus
  0.4em\relax IEEE, 2013, pp. 6645--6649.

\bibitem{peddinti2015time}
V.~Peddinti, D.~Povey, and S.~Khudanpur, ``A time delay neural network
  architecture for efficient modeling of long temporal contexts,'' in
  \emph{Sixteenth Annual Conference of the International Speech Communication
  Association}, 2015.

\bibitem{povey2011kaldi}
D.~Povey, A.~Ghoshal, G.~Boulianne, L.~Burget, O.~Glembek, N.~Goel,
  M.~Hannemann, P.~Motlicek, Y.~Qian, P.~Schwarz \emph{et~al.}, ``The kaldi
  speech recognition toolkit,'' in \emph{IEEE 2011 workshop on automatic speech
  recognition and understanding}, no. CONF.\hskip 1em plus 0.5em minus
  0.4em\relax IEEE Signal Processing Society, 2011.

\bibitem{povey2016purely}
D.~Povey, V.~Peddinti, D.~Galvez, P.~Ghahremani, V.~Manohar, X.~Na, Y.~Wang,
  and S.~Khudanpur, ``Purely sequence-trained neural networks for asr based on
  lattice-free mmi.'' in \emph{Interspeech}, 2016, pp. 2751--2755.

\bibitem{jacob2018quantization}
B.~Jacob, S.~Kligys, B.~Chen, M.~Zhu, M.~Tang, A.~Howard, H.~Adam, and
  D.~Kalenichenko, ``Quantization and training of neural networks for efficient
  integer-arithmetic-only inference,'' in \emph{Proceedings of the IEEE
  Conference on Computer Vision and Pattern Recognition}, 2018, pp. 2704--2713.

\bibitem{povey2018semi}
D.~Povey, G.~Cheng, Y.~Wang, K.~Li, H.~Xu, M.~Yarmohammadi, and S.~Khudanpur,
  ``Semi-orthogonal low-rank matrix factorization for deep neural networks.''
  in \emph{Interspeech}, 2018, pp. 3743--3747.

\bibitem{paszke2019pytorch}
A.~Paszke, S.~Gross, F.~Massa, A.~Lerer, J.~Bradbury, G.~Chanan, T.~Killeen,
  Z.~Lin, N.~Gimelshein, L.~Antiga \emph{et~al.}, ``Pytorch: An imperative
  style, high-performance deep learning library,'' in \emph{Advances in Neural
  Information Processing Systems}, 2019, pp. 8024--8035.

\bibitem{wong2016sequence}
J.~H. Wong and M.~Gales, ``Sequence student-teacher training of deep neural
  networks,'' in \emph{Proc. of Interspeech 2016}, 2016.

\bibitem{keith2018optimizing}
F.~Keith, W.~Hartmann, M.-H. Siu, J.~Ma, and O.~Kimball, ``Optimizing
  multilingual knowledge transfer for time-delay neural networks with low-rank
  factorization,'' in \emph{2018 IEEE International Conference on Acoustics,
  Speech and Signal Processing (ICASSP)}.\hskip 1em plus 0.5em minus
  0.4em\relax IEEE, 2018, pp. 4924--4928.

\bibitem{krishnamoorthi2018quantizing}
R.~Krishnamoorthi, ``Quantizing deep convolutional networks for efficient
  inference: A whitepaper,'' \emph{arXiv preprint arXiv:1806.08342}, 2018.

\bibitem{xue2013restructuring}
J.~Xue, J.~Li, and Y.~Gong, ``Restructuring of deep neural network acoustic
  models with singular value decomposition.'' in \emph{Interspeech}, 2013, pp.
  2365--2369.

\bibitem{abadi2016tensorflow}
M.~Abadi, P.~Barham, J.~Chen, Z.~Chen, A.~Davis, J.~Dean, M.~Devin,
  S.~Ghemawat, G.~Irving, M.~Isard \emph{et~al.}, ``Tensorflow: A system for
  large-scale machine learning,'' in \emph{12th $\{$USENIX$\}$ Symposium on
  Operating Systems Design and Implementation ($\{$OSDI$\}$ 16)}, 2016, pp.
  265--283.

\bibitem{can2018pykaldi}
D.~Can, V.~R. Martinez, P.~Papadopoulos, and S.~S. Narayanan, ``Pykaldi: A
  python wrapper for kaldi,'' in \emph{2018 IEEE International Conference on
  Acoustics, Speech and Signal Processing (ICASSP)}.\hskip 1em plus 0.5em minus
  0.4em\relax IEEE, 2018, pp. 5889--5893.

\bibitem{madikeri2020pkwrap}
S.~Madikeri, S.~Tong, J.~Zuluaga-Gomez, A.~Vyas, P.~Motlicek, and H.~Bourlard,
  ``Pkwrap: a pytorch package for lf-mmi training of acoustic models,''
  \emph{arXiv preprint arXiv:2010.03466}, 2020.

\bibitem{hadian2018end}
H.~Hadian, H.~Sameti, D.~Povey, and S.~Khudanpur, ``End-to-end speech
  recognition using lattice-free mmi.'' in \emph{Interspeech}, 2018, pp.
  12--16.

\bibitem{panayotov2015librispeech}
V.~Panayotov, G.~Chen, D.~Povey, and S.~Khudanpur, ``Librispeech: an asr corpus
  based on public domain audio books,'' in \emph{2015 IEEE International
  Conference on Acoustics, Speech and Signal Processing (ICASSP)}.\hskip 1em
  plus 0.5em minus 0.4em\relax IEEE, 2015, pp. 5206--5210.

\end{thebibliography}

\end{document}